# Photoacoustic microscopical simulation platform for large volumetric imaging using Bessel beam

Xianlin Song [a, #, *], Jianshuang Wei [b, c, #], Lingfang Song [d]

[a] School of Information Engineering, Nanchang University, Nanchang, China;
[b] Britton Chance Center for Biomedical Photonics, Wuhan National Laboratory for Optoelectronics-Huazhong University of Science and Technology, Wuhan, China;
[c] Moe Key Laboratory of Biomedical Photonics of Ministry of Education, Department of Biomedical Engineering, Huazhong University of Science and Technology, Wuhan 430074, China;
[d] Nanchang Normal University, Nanchang 330031, China;
[#] equally contributed to this work
*songxianlin@ncu.edu.cn

**Abstract**：We developed a Bessel-beam photoacoustic microscopical simulation platform by using the k-Wave: MATLAB toolbox. The simulation platform uses the ring slit method to generate Bessel beam. By controlling the inner and outer radius of the ring slit, the depth-of-field (DoF) of Bessel beam can be controlled. And the large volumetric image is obtained by point scanning. The simulation experiments on blood vessels was carried out to demonstrate the feasibility of the simulation plat-form. This simulation work can be used as an auxiliary tool for the research of Bessel-beam photoacoustic microscopy.

## 1 Introduction

Photoacoustic imaging is a promising technique that combines optical contrast with ul-trasonic detection to map the distribution of the absorbing pigments in biological tissues [1]-[4]. It has been widely used in biological researches, such as structural imaging of vasculature [5], brain structural and functional imaging [6], and tumor detection [7]. Considering the lateral resolution of photoacoustic microscopy (PAM), it can be classi-fied into two categories: optical-resolution (OR-) and acoustic-resolution (AR-) PAM [8, 9]. In AR-PAM, the spatial resolution is determined by the acoustic focus, since the laser light is weekly or even not focused on the sample. Conversely, in the OR-PAM, the laser light is tightly focused into the sample to achieve sharp excitation. However, in OR-PAM,  the DoF is quite limited, for it is determined by the optical condenser and is closely relat-ed to the optical focusing. The small DoF will prevent OR-PAM to achieve high-quality 3D images or acquire dynamic information in depth direction.

Depth scanning using motorized stage is widely used to address this issue as it is the most convenient method [10] [11]. To avoid slow mechanical scanning in the depth di-rection, several methods by engineering the illumination have been proposed. Double il-lumination can double the DoF by illuminating the sample from both top and bottom sides simultaneously. But it is only valid in transmission-mode OR-PAM [12]. Utilizing chromatic aberration of non-achromatic objective, multi-wavelength laser can generate multi-focus along the depth direction [13]. However, this method sacrifices the capability of functional imaging. Electrically tunable lens (ETL) has also been introduced in OR-PAM [14], the focal plane can be settled in about 15 ms (EL-10-30，Optotune AG). It is fast enough for pulsed lasers with a repetition rate of hundred-hertz, while being quite slow for those lasers with repetition rate of hundreds of kilo-hertz.

Non-diffraction beam inherently own a large DoF. Photoacoustic Microscopy Based on Bessel Beam can achieve extended DoF with retaining high lateral resolution [15]. In this paper, we developed a Bessel beam photoacoustic microscopical simulation plat-form with extened DoF by using the k-Wave:

MATLAB toolbox. The simulation plat-form uses the ring slit method to generate Bessel beam. By controlling the inner and outer radius of the ring slit, the DoF of Bessel beam can be controlled. And the large volumetric image is obtained by point scanning. And the simulation experiments on blood vessels was carried out to demonstrate the feasibility of simulation platform.

## 2 Simulation method

### 2.1 Configuration Environment

The k-Wave simulation toolbox can analyze photoacoustic signals in the time domain [16]. We use k-Wave: MATLAB toolbox for the simulation of Bessel-beam photoacous-tic microscopy. The simulated environment was created in three dimension with 100 x 100 x 100 voxels (each voxel size is 2 μm), as shown in Figure 1, and contains a perfectly matched boundary layer (PML) to satisfy the boundary conditions for the forward pro-cess. The surrounding medium is water with a sound velocity of 1.5 km/s and a density of 1000 kg/m3. All simulations assume that an acoustically homogeneous medium was considered with no absorption or dispersion of sound.

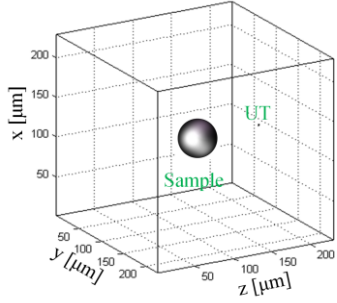

Figure 1. Schematic diagram of simulation of Bessel-beam photoacoustic microscopy.

In our simulation, the parameters of the simulated system were chosen to match those of the practical setup as closely as possible, the center frequency of the ultrasonic transducer is set to 50 MHz and the bandwidth is 80%. The three-dimensional imaging data can be obtained by carried out two-dimensional raster scan.

### 2.2 Generation of Bessel beams

Based on Fourier optics theory, since the Fourier transform of a ring is a zero-order Bessel function, placing a ring slit on the back focal plane of the lens can convert incident light into a Bessel beam. If the annular slit is symmetrical about the optical axis, the field intensity distribution near the front focal plane can be known from the scalar diffraction theory

$$E(r,z) = A\frac{e^{i\kappa z}}{i\lambda f} \int P(\rho) J_0(\frac{\kappa}{f} r\rho) e^{-i\frac{\rho^2}{f^2}\kappa z} \rho d\rho \quad (1)$$

Where $P(\rho)$ is the pupil function, $f$ is the lens focal length, $\lambda$ is the wavelength, and $A$ is a constant representing the amplitude. $P(\rho)$ is a circular slit and substituting it into the above formula, the distribution of the approximate Bessel function can be obtained. The field distribution of the slit with an outer ring radius of 5 mm near the focal point of the objective lens with a focal length of 40 mm is shown in Figures 2. Figure 2 (a) is the Bessel beam generating from an annular slit with an outer ring radius of 5 mm and a slit width of 150 μm. Figure 2 (b) is the Bessel beam generating from an annular slit with an outer ring radius of 5 mm and a slit width of 400 μm. As the slit width increases, the DoF decreases. Figure 2 (c) is the intensity distribution in focal plane indicated by the yellow dashed line in Figure 2 (b). The lateral resolution can be calculated from the full height at half maximum (FWHM) of the intensity distribution of Bessel beam in focal plane, and the FWHM is ~ 1.2 μm.

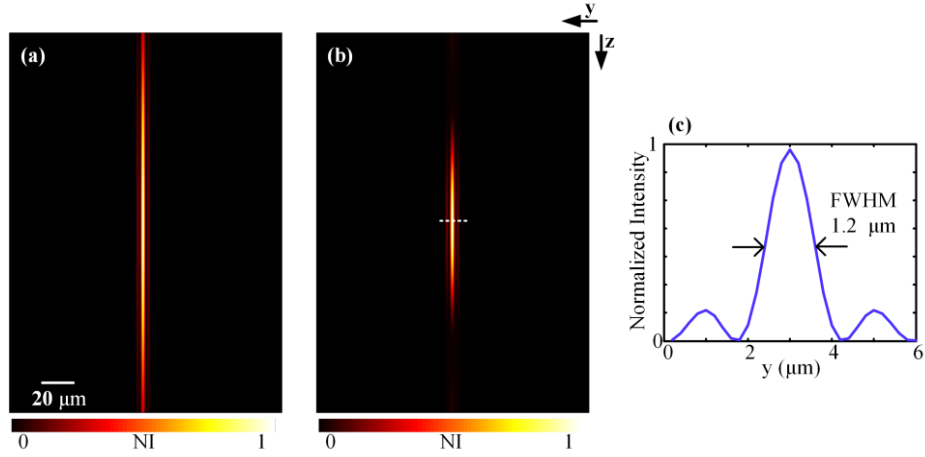

Figure 2. Bessel beam generating from an annular slit with different slit width. (a) Bessel beam generating from an annular slit with an outer ring radius of 5 mm and a slit width of 150 μm. (b) Bessel beam generating from an annular slit with an outer ring radius of 5 mm and a slit width of 400 μm. (c) Intensity distribution in focal plane indicated by the yellow dashed line in (b). Objective focal length 40 mm, wavelength 582 nm. NI, normalized intensity.

## 3 RESULTS

### 3.1 Imaging of the vasculature

A virtual vascular network is used to verify the imaging capabilities of the constructed simulation platform, as shown in Figure 3(a). The diameter of the blood vessel is 2 - 10 μm. The blood vessel was placed at a depth of about 300 μm from the ultrasonic transducer, and a two-dimensional raster scan was performed with a step size of 2 μm to obtain three-dimensional imaging data. Figure 3(b) is MAP image. Figures 3(c) and 3(d) are the close-up images of the small areas indicated by the white dashed rectangles in Figures 3(a) and 3(b), respectively. The structure of blood vessels can be distinguished, but the image appears blurred due to the side lobe of Bessel beams (indicated by the white arrows in Figure 3(d)). To quantitatively demonstrate the performance of our simulation platform, the photoacoustic signal distribution of a vessel (indicated by the white dashed line in Figures 3(a) and 3(b)) was chosen for width analysis, the corresponding width of the vessel was defined as the FWHM of the photoacoustic signal curve. As shown in Figure 3(e), the width of original vessel is ~2 μm, while, since the side lobe of Bessel beams can deteriorate the resolution, the width of vessel is measured to be ~9.5 μm.

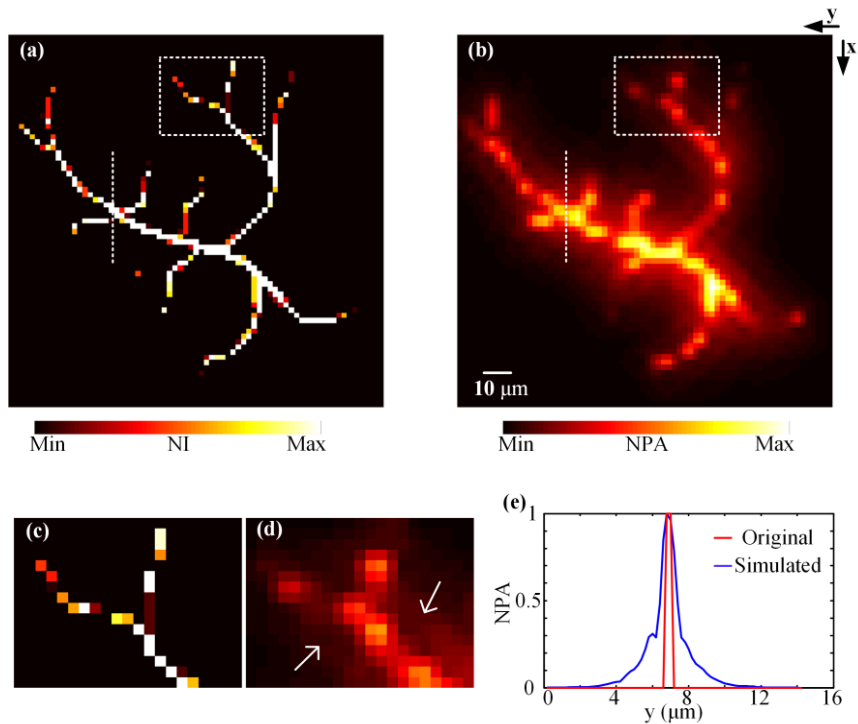

Figure 3. Images of vasculature. (a) Original blood vessels for simulation. (b) The MAP images of vasculature. (c) and (d) are the close-up images of the small areas indicated by the white dashed rectangles in (a) and (b), respectively. (e) is the photoacoustic signal distribution of a vessel (indicated by the white dashed line in (a) and (b)). NPA, normalized photoacoustic amplitude.

## 4 Conclusion

We have developed a Bessel-beam photoacoustic microscopical simulation platform using k-Wave simulation toolbox. Bessel beam was generated from annular slit. And the changes in slit width cause changes in DoF, larger slit width generated narrow DoF. The slit width can be set as experimental requirement. Two-dimensional raster scan was per-formed to obtained volumetric data. A virtual vascular network is used to verify the im-aging capabilities of the constructed simulation platform. This work will contribute to the study of Bessel-beam photoacoustic microscopy.